\documentclass[aps,twocolumn,prb,superscriptaddress]{revtex4-1}
\usepackage[utf8x]{inputenc}
\usepackage{amsmath,amssymb,graphicx,xcolor}
\newlength{\mylength}
\renewcommand\Re{\operatorname{Re}}

\newcommand{\Parentesis}{\left( \ldots \psi_{k-1}, \psi_{k+1} \ldots \right)}

\newcommand{\Ezero}{ {\bf e}^{\left(0\right)}}	
	
\newcommand{\Einczero}{ {\bf e}_{inc}^{\left( 0 \right)}}	
\newcommand{\Eincone}{ {\bf e}_{inc}^{\left( 1 \right)}}	
\newcommand{\Einctwo}{ {\bf e}_{inc}^{\left( 2 \right)}}

\newcommand{\Eone}{ {\bf e}^{\left(1\right)}}
\newcommand{\Etwo}{ {\bf e}^{\left(2\right)}}

\newcommand{\tEtwo}{ \bar{\bf e}^{\left(2\right)}}
\newcommand{\ttEtwo}{ {\bf \bar{\bar{e}}}^{\left(2\right)}}

\newcommand{\Hzero}{ {\bf h}^{\left(0\right)}}

\newcommand{\Hone}{ {\bf h}^{\left(1\right)}}

\newcommand{\Htwo}{ {\bf h}^{\left(2\right)}}
\newcommand{\chizero}{ {\chi}_{\mathtt{SC}}^{\left(0\right)}}
\newcommand{\lambdaSC}{\lambda_{\mathtt{SC}}}

\newcommand{\epsr}{ {\varepsilon}_{\mathtt{SC},r}}
\newcommand{\epszeror}{ {\varepsilon}_{\mathtt{SC},r}^{\left(0\right)}}
\newcommand{\epsoner}{ {\varepsilon}_{\mathtt{SC},r}^{\left(1\right)}}
\newcommand{\epstwor}{ {\varepsilon}_{\mathtt{SC},r}^{\left(2\right)}}

\newcommand{\rkstwz}{r_{s,k}''^{\left(0\right)}}	

\newcommand{\rkinctwz}{r_{inc,k}''^{\left(0\right)}}	
\newcommand{\rkinconz}{r_{inc,k}'^{\left(0\right)}}
\newcommand{\rkinctwo}{r_{inc,k}''^{\left(1\right)}}	
\newcommand{\rkincono}{r_{inc,k}'^{\left(1\right)}}
\newcommand{\rkinctwt}{r_{inc,k}''^{\left(2\right)}}	
\newcommand{\rkincont}{r_{inc,k}'^{\left(2\right)}}
\newcommand{\rkstwo}{r_{s,k}''^{\left(1\right)}}	

\newcommand{\rkstwt}{\tilde{r}_{s,k}''^{\left(2\right)}}	
\newcommand{\rksont}{\tilde{r}_{s,k}'^{\left(2\right)}}
\newcommand{\domain}{\quad \forall t \in \left\{ 1, 2, 3 \right\}}

\newcommand{\sigmaon}{\sigma'}
\newcommand{\sigmatw}{\sigma''}

\newcommand{\sigmaonk}{\sigma_{k}'}

\newcommand{\sigmatwk}{\sigma_{k}''}
\newcommand{\PkAlpha}{p_{k, \hat{\alpha}}}
\newcommand{\tauon}{\tau'}
\newcommand{\tautw}{\tau''}
\newcommand{\tauonh}{\tau_{h}'}
\newcommand{\tauonk}{\tau_{k}'}
\newcommand{\tautwh}{\tau_{h}''}
\newcommand{\tautwk}{\tau_{k}''}
\newcommand{\Einc}{{\bf E}_{inc}}

\begin{document}
\title{Cloaking of Arbitrarily-Shaped Objects with Homogeneous Coatings}
\author{Carlo Forestiere}\email{Corresponding author: carlo.forestiere@gmail.com}
\affiliation{Department of Electrical and Computer Engineering \& Photonics Center, Boston University, 8 Saint Mary’s Street, Boston, Massachusetts}
\affiliation{ Department of Electrical Engineering and Information Technology, Universit\`{a} degli Studi di Napoli Federico II, via Claudio 21,
 Napoli, 80125, Italy}
\author{Luca Dal Negro}
\affiliation{Department of Electrical and Computer Engineering \& Photonics Center, Boston University, 8 Saint Mary’s Street, Boston, Massachusetts}
\author{Giovanni Miano}
\affiliation{ Department of Electrical Engineering and Information Technology, Universit\`{a} degli Studi di Napoli Federico II, via Claudio 21,
 Napoli, 80125, Italy}

%\section{Formulation}
%\
\begin{abstract}
We present a theory for the cloaking of arbitrarily-shaped objects and demonstrate electromagnetic scattering-cancellation through designed
homogeneous coatings. First, in the small-particle limit, we expand the dipole moment of a coated object in terms of its resonant modes. By
zeroing the numerator of the resulting rational function, we accurately predict the permittivity values of the coating layer that abates the total
scattered power. Then, we extend the applicability of the method beyond the small-particle limit, deriving the radiation corrections of
the
scattering-cancellation permittivity within a perturbation approach. Our method  permits the design of “invisibility cloaks” for irregularly-shaped
devices such as complex sensors and detectors.
\end{abstract}
\maketitle
%% WC 492

Arguably the most studied inverse scattering problem, the design of a cloaking environment  that drastically reduces or ideally cancels
the electromagnetic scattering of a given object,  fascinates scientists and engineers from a broad  variety  of disciplines. The most popular
cloaking approaches developed in the past decade are based on transformation-optics
\cite{Pendry06,Leonhardt06} and scattering-cancellation \cite{Alu05,Silveirinha08} and have both been experimentally
validated at microwaves \cite{Schurig06,Rainwater12}. In particular, the latter technique presents several practical advantages, since
it only makes use of homogeneous and isotropic materials, which are easier to fabricate than the inhomogeneous and anisotropic media required by the
transformation-optics approach. In addition, the interior of the cloaked object supports non-vanishing fields: this fact has inspired the seminal idea
of a cloaked-sensor which can drastically reduce the perturbation introduced by the measuring apparatus on the physical quantity under investigation
\cite{Alu09}. However,  a rigorous solution of the scattering-cancellation problem exists only for simple shapes, such as cylinders or spheres 
due to the substantial difficulties in tackling the inverse scattering problem in the presence of non-canonical geometries.  This fact may
prevent the application of the scattering-cancellation approach to real-life problems since the shape of the object to be cloaked is in general not
under the control of the designer of the cloaking system.
So far, this problem has been circumvented by  resorting to numerical optimization techniques \cite{Silveirinha08,Tricarico2010}. Unfortunately,
besides the lack of physical understanding, numerical optimization usually presents a high computational burden, requiring a large number of
iterations of the direct electromagnetic problem. In addition, it is also well-known that numerical optimization techniques can be trapped in local
minima. This fact is particularly relevant to the problem at hand due to the existence of multiple solution of different quality.  Deriving general
design techniques for the electromagnetic properties of the homogeneous coating of an arbitrarily shaped
object
to achieve scattering-cancellation remains a grand challenge.

 In this paper, we address this problem introducing a general theory of
scattering-cancellation from an arbitrarily shaped object with a homogeneous coating. Our theory enables the rigorous design of
the permittivity of the cover of an arbitrarily shaped object to achieve cloaking or ``invisibility'' in the limit of low-losses. 
The problem is tackled in two-steps. First, in the small-particle limit (Rayleigh regime), we expand the dipole moment of the coated object in
terms of its electrostatic modes, using the theoretical framework developed in Refs.
\cite{Ouyang89,Fredkin03,Mayergoyz05,Mayergoyz07,ForestierePRB_13}.
By noting that the obtained expansion is a rational function of the dielectric permittivity of the coating, we can determine the
scattering-cancellation condition by zeroing its numerator. Next, in order to extend our method beyond the Rayleigh regime, we derive the radiation
corrections of the scattering-cancellation permittivity by using the perturbation approach introduced by Mayergoyz et al. in Ref.
\cite{Mayergoyz05,Hung13}. In particular, the first- and second-order
radiation corrections of the scattering-cancellation permittivity are found by zeroing the corresponding-order perturbation of the dipole
moment.  Then, we validate our method and estimate its accuracy by designing the
susceptibility of the cover of a sphere of several electric sizes. Finally, we design the invisibility cloaking of a C-shaped particle, showing
that, by using the radiation corrections, the scattering-cancellation permittivity can be accurately predicted even for objects of size
comparable to
the incident wavelength.

\section{SMALL-PARTICLE LIMIT}

\begin{figure}[ht]
\centering
\includegraphics[width=\mylength]{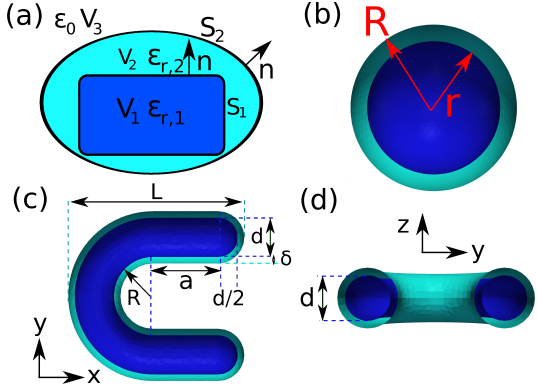}
\caption{Sketch of the studied core-shell geometry (a). Investigated spherical shell (b). Investigated coated C-shaped object where only the
portion
of the boundary surfaces with $z<0$ (c) and $x<0$ (d) is shown. }
  \label{fig:Figure_1}
\end{figure}

We start by considering a core-shell object of arbitrary shape sketched in Fig. \ref{fig:Figure_1} (a), and embedded in free-space. The core is
assumed to be made of a linear, homogeneous, isotropic, lossless medium with relative permittivity
$\varepsilon_{r,1} \in \mathbb{R}$ and corresponding susceptibility $\chi_1 = \varepsilon_{r,1} - 1$, whereas the shell is composed by a linear,
homogeneous,
isotropic, and time-dispersive material exhibiting a complex permittivity $\varepsilon_{r,2} \left( \omega \right) = \varepsilon_{r,2}' - j
\varepsilon_{r,2}''$ (a time-harmonic dependence $e^{j \omega t }$ has been assumed) and a corresponding susceptibility $\chi_2\left( \omega
\right)= \varepsilon_{r,2} - 1$. We denote with $V_1$ and $V_2$ the volumes occupied by the core and the shell, respectively, and with $V_3$ the
external
space. We also denote
with $S_1$ and $S_2$ the surfaces separating the shell with the core and with the external space, respectively. Both the
outward-pointing normals to the two surfaces $S_1$ and $S_2$ are indicated with $\bf n$. Assuming that the investigated system is much smaller than
the
wavelength of operation we employ the quasi-electrostatic approximation of the Maxwell's equations.
%%%%%%%%%%%%%%%%%
%%%%%%%%%%%%%%%%%
%% WC 55

The source-free electric field that may exist in the presence of a dielectric coating with $\varepsilon_{r,2}' < 0$ can be  described by two
equivalent free-standing single layers of electric charge density $\sigmaon$ and $\sigmatw$ distributed on $S_1$ and
$S_2$, respectively. They are solution of the following homogeneous boundary integral equation \cite{Mayergoyz07}:
\begin{equation}
  \mathcal{A}
\begin{bmatrix}
\sigmaon   \\
\sigmatw   
\end{bmatrix}
= \\
\beta \,
\mathcal{B}
\begin{bmatrix}
\sigmaon   \\
\sigmatw   
\end{bmatrix}, 
\label{eq:EigProblem_Sigma}
\end{equation}
%% WC 16
where $\mathcal{A}$ and $\mathcal{B}$ are endomorphisms of the vector space $ \mathbb{L}^2 \left( S_1 \right) \times \mathbb{L}^2 \left( S_2 \right)$
defined as:
\begin{align}
\mathcal{A} = &
\begin{bmatrix}
\varepsilon_{r,1} \left( \mathcal{L}_{11} - \mathcal{I} \right) &  \varepsilon_{r,1} \mathcal{L}_{12}   \\
        -\mathcal{L}_{21}  &   \left( -\mathcal{L}_{22} - \mathcal{I} \right)
\end{bmatrix}, 
\\
\mathcal{B}  = &
\begin{bmatrix}
 \left( \mathcal{L}_{11} + \mathcal{I} \right) & \mathcal{L}_{12}   \\
 -\mathcal{L}_{21}  &   \left( -\mathcal{L}_{22} + \mathcal{I} \right)
\end{bmatrix}, 
\end{align}
%% WC 15
$ \mathcal{L}_{uv}: \mathbb{L}^2 \left( S_v \right) \rightarrow \mathbb{L}^2 \left(
S_u	\right)$   is defined $\forall u,v \in \left\{1,2\right\}$ as 
\begin{equation}
 \mathcal{L}_{uv} \left\{  \sigma \right\}\left(Q \right)  =   \frac{1}{2\pi} \oint_{S_v} \sigma \left( M \right) \frac{{\bf r}_{MQ}
\cdot {\bf n}_Q
}{{r}_{MQ}^3} dS_M, \, Q \in S_u
 \label{eq:Operator_L}
\end{equation}
%% WC 62
and $\mathcal{I}$ is the identity operator.
Equation \ref{eq:EigProblem_Sigma} defines a generalized eigenvalue
problem in $\beta$ for the operator pair $\left( \mathcal{A}, \mathcal{B} \right)$.
As the operators $\mathcal{A}$ and $\mathcal{B}$ are not self-adjoint, the eigenmodes of the problem (1) are not orthogonal. Therefore, we
consider the following problem:
\begin{equation}
  \mathcal{A}^\dagger
\begin{bmatrix}
\tauon   \\
\tautw  
\end{bmatrix}
= \\
\beta \,
\mathcal{B}^\dagger
\begin{bmatrix}
\tauon   \\
\tautw  
\end{bmatrix}
\label{eq:EigProblem_Tau},
\end{equation}
%% WC 55
where $\mathcal{A}\dagger$ and $\mathcal{B}\dagger$ are the adjoint operators of $ \mathcal{A}$ and $\mathcal{B}$. Since the operators $ \mathcal{A}$
and $\mathcal{B}$ are compact, problems \ref{eq:EigProblem_Sigma} and \ref{eq:EigProblem_Tau} support discrete
spectra. Moreover, problems \ref{eq:EigProblem_Sigma} and \ref{eq:EigProblem_Tau} share the same eigenvalues
$\left\{ \beta_k | k \in \mathbb{N} \right\}$, whereas their eigenmodes $\left\{ \sigma_k|  k \in \mathbb{N} \right\}$ and $\left\{ \tau_k|  k \in
\mathbb{N} \right\}$ form bi-orthonormal sets \cite{Mayergoyz07}, namely:
\begin{multline}
 \oint_{S_1} \tauonh \left( \mathcal{L}_{11} + \mathcal{I} \right) 
\sigmaonk dS \\
+ \oint_{S_1}\tauonh \, \mathcal{L}_{12} \sigmatwk \, dS 
- \oint_{S_2} \tautwh \, \mathcal{L}_{21} \sigmaonk \, dS  \\
- \oint_{S_2} \tautwh  \left( \mathcal{L}_{22} - \mathcal{I} \right) \sigmatwk dS = \delta_{hk},
\label{eq:Nomalization}
\end{multline}
%%% WC 94
where $\delta_{h,k}$ is the Kronecker delta, which is 1 if the $k=h$, and 0 otherwise.  It is fundamental to note that the eigenvalues
$\beta_k$, and the associated eigenmodes depend solely on the geometry of the shell and on the dielectric constant of the core. This  fact  is crucial
for the design of the dielectric constant of the cover to achieve transparency. It can be also proved  that the eigenvalue $\beta_k$ are real and  
negative. 

We associate to each eigenmode $k$ a dipole moment ${\bf p}_k = \left(p_{k,x}, p_{k,y},p_{k,z} \right)$
\begin{equation}
   {\bf p}_k = \oint_{S_1} {\bf r} \sigmaonk  dS + \oint_{S_2} {\bf r} \sigmatwk  dS. 
   \label{eq:DipolarMomentMode}
\end{equation}
%%% WC  14
%% WC 36
The kth plasmonic mode is {\it dark} if  ${\bf p}_k = 0$, it is {\it bright} otherwise.

When the coated object is excited by an external field ${\bf E}_{inc}$ its dipole moment can be expressed as
\cite{Mayergoyz05,ForestierePRB_13}:
 \begin{equation}
     \label{eq:DipolarTotal}
     {\bf p}^{\left( 0 \right)} \left( \chi_2 \right) =  \displaystyle\sum_k \frac{ \chi_1 \, \rkinconz+ \chi_{2} \rkinctwz}{ \psi_k - \chi_2 }{\bf
p}_k,
 \end{equation}
where:
\begin{equation}
 \begin{aligned}
  \rkinconz  &= - {2\varepsilon_0}  \oint_{S_1} {\bf E}_{inc} \cdot {\bf n} \tauonk dS,  \\ 
  \rkinctwz  &=   {2\varepsilon_0} \left( \oint_{S_1} {\bf E}_{inc} \cdot {\bf n} \tauonk dS  -
\oint_{S_2} {\bf E}_{inc} \cdot {\bf n} \tautwk dS \right) .
  \end{aligned}  
   \label{eq:rk_zero_order}
\end{equation}
%% WC 63
The  real resonant frequency $\omega_{k}$ of the mode  $k$ can be obtained by the equation:
\begin{equation}
 \Re \left\{ \varepsilon_{r,2}  \left( \omega_k \right) \right\} = \beta_k,
 \label{eq:ResonantFrequency}
\end{equation}
Equation \ref{eq:DipolarTotal} shows that the  total dipole moment is a rational function of the susceptibility of the cover $\chi_2$, since as
already noticed $\psi_k$, $\rkinconz$, $\rkinctwz$, and ${\bf p}_k$ are independent of $\chi_2$. This fact allows a simple and elegant solution of the
inverse 
scattering problem.

The quantity $c_k \left( \chi_2 \right) = \left[ \chi_1 \, \rkinconz + \chi_{2} \rkinctwz \right]$ in
the numerator of Eq. \ref{eq:DipolarTotal} represents the coupling coefficient of the mode $k$ to the external excitation $\Einc$.  Moreover, we
define the resonant radiative strength ${\bf s}_{k}$ of the mode $k$ 
\begin{equation}
  {\bf s}_{k}= c_k \left( \psi_k \right) {\bf p}_k  = \left[ \chi_1 \, \rkinconz + \psi_k \rkinctwz \right]  {\bf
p}_k.
\label{eq:RadiativeStrenght}
\end{equation}
%% WC 65
which quantifies the contribution of the mode $k$ to the total dipole moment at its resonance. It is immediate to notice
that dark modes have vanishing magnitude of ${\bf s}_{k}$. As we will see later by examples, this synthetic parameter is particularly
useful to separate the modes that make a significant contribution to the dipole moment by those who make a minor contribution and can therefore be
neglected. 

%% WC 17
Once the total dipole moment is known, the total power scattered by the structure is given by:
\begin{equation} 
  P_{rad} = \frac{\omega^4} { 12 \pi \varepsilon_0 c^3} \left| {\bf p} \right|^2 =  \frac{\omega^4} { 12 \pi \varepsilon_0 c^3}
\displaystyle\sum_{t\in \left\{ x, y, z \right\}}  \left| { p}_t \right|^2,
  \label{eq:RadiatedPower}
\end{equation}
%% WC 9
being $c$ the speed of light in free-space. 
%% WC 150

The goal of our study is to find the values of susceptibility of the cover at which the scattered power $P_{rad}$ vanishes when the
core-shell object is excited by the field ${\bf E}_{inc}$. We assume that the coated object exhibits $n$ bright modes, i.e.
$\sigma_k = \left( \sigmaonk, \sigmatwk \right) \, | \, k=1 \ldots n  $, with corresponding resonant
susceptibilities $\psi_k$. 
We also assume that the component of the total dipole moment along a given direction $  \hat{\boldsymbol \alpha}$ is strongly dominant in the
frequency range of interest:
\begin{equation}
\frac{\left| { {\bf p} \cdot \hat{\boldsymbol \alpha}} \right|}{ \left\| {\bf p} - \left( {\bf p} \cdot \hat{\boldsymbol \alpha} \right)
\hat{\boldsymbol \alpha} \right\|
} \gg 1.
\label{eq:Hypothesis}
\end{equation}
Thus, the problem reduces to finding the values of $\chi_2$ in correspondence of which the  $\hat{\boldsymbol \alpha}$-component of the dipole moment
vanishes. In the presence of objects of moderate aspect ratio, the offset between the directions of the dipole moment and the incident polarization
direction is usually small, thus it is reasonable to assume $\hat{\boldsymbol \alpha} =  \Einc /{\left\|  \Einc \right\|}$.
Starting from Eq. \ref{eq:DipolarTotal} it is straightforward to demonstrate that the zeros of the  $ \hat{\boldsymbol \alpha}$ component of
the dipole moment are given by the roots of the following polynomial of degree $n$
\begin{multline}
 \mathcal{P}_{\hat{ \alpha}} \left\{ \chi_2 \right\} = 
  \chi_2^{ n }    \displaystyle\sum_{k=1}^n \rkinctwz \PkAlpha  \\
 -  \chi_2^{ n -1} \displaystyle\sum_{k=1}^n \left[\rkinctwz e_1 \Parentesis  - \chi_1  \rkinconz  \right] \PkAlpha  + \\
 + \chi_2^{ n -2} \displaystyle\sum_{k=1}^n 
  \left[  \rkinctwz e_2 \Parentesis \right. \\
 \left. - \chi_1 \rkinconz e_1 \Parentesis \right] \PkAlpha  + \cdots = 0,
\label{eq:DesignFormula}
\end{multline}
%%% WC 29
where $ \PkAlpha = {\bf p}_k \cdot   \hat{\alpha}$ and $e_i \Parentesis $ is an {\it elementary symmetric
polynomial} of degree $i$ in the $n-1$
variables $\psi_1, \ldots, \psi_{k-1}, \psi_{k+1}, \ldots, \psi_{n}$, defined as  \cite{Borwein95}:
\begin{equation}
\label{eq:SimmetricPoly}
\begin{aligned}
  e_0  \Parentesis &= 1,\\
  e_1 \Parentesis  &= \displaystyle\sum_{\substack{1 \leq l \leq n \\ l \ne k }} \psi_l,\\
  e_2 \Parentesis &= \displaystyle\sum_{\substack{1 \leq l < u \leq n \\ l, u \ne k }} \psi_l
\psi_u,\\
  e_3 \Parentesis &=  \displaystyle\sum_{\substack{1 \leq l < u< v \leq n \\ l, u, v \ne k }}
\psi_l
\psi_u \psi_v,\\
 \ldots
\end{aligned}
\end{equation}
%% WC 31
 Equation \ref{eq:DesignFormula} is indeed a design formula: denoting with $\chizero$ the generic root of the polynomial
$\mathcal{P}_{\hat{ \alpha}}$, to
achieve transparency at a
prescribed frequency $\omega_0$ the  cover has to exhibit a susceptibility $  \chi_2 \left( \omega_0 \right) = \chizero$. We also
denote with $\epszeror = \chizero + 1$ the corresponding value of scattering-cancellation permittivity. It should be emphasized that at
microwaves or in the far infrared, a metamaterial cover exhibiting the prescribed  susceptibility can be engineered and fabricated (e.g.
\cite{EngetaBook}). Instead, in the visible, the difficulties
in manipulating strongly sub-wavelength meta-atoms make the implementation of the
designed cover very challenging.

 \section{Radiation correction of the scattering-cancellation permittivity}
As the diameter $D$ of the minimum sphere circumscribing the particle becomes comparable to the incident wavelength $\lambda$, the quasi-static
prediction made by Eq. \ref{eq:DesignFormula} becomes inaccurate. Nevertheless, our method can be significantly extended by using a perturbation
approach.  We have adapted to the problem at hand the approach that  Mayergoyz et al. originally introduced to study the radiation correction of the
plasmonic resonance \cite{Mayergoyz05}. By introducing the perturbation parameter $\beta = \omega \sqrt{\epsilon_0 \mu_0} D$, we expand the
excitation fields, the relative permittivity ${\varepsilon_{\mathtt{SC},r}}$ at which the scattering-cancellation occurs, and the corresponding
electric and magnetic fields in powers of $\beta$.  In particular, for ${\varepsilon_{\mathtt{SC},r}}$ we have
\begin{equation}
   {\varepsilon_{\mathtt{SC},r}}  \approx {\epszeror} + \beta \epsoner +\beta^2 \epstwor + \cdots.
\end{equation}
These expansions are then substituted into the Maxwell's equations and first- and second-order boundary value problems are obtained by equating the
terms of corresponding order. Eventually, the first- and second-order radiation corrections of the scattering-cancellation permittivity, i.e.
$\epsoner$ and $\epstwor$, are found by zeroing the corresponding perturbations of the $\hat{\boldsymbol \alpha}$-component of the dipole moment of
the coated object. The detailed derivation is shown in the appendix. As we will see in the next section, the quasi-electrostatic theory accompanied by
the radiation corrections turns out to be the most natural and predictive framework to  describe the regime in which the dipole scattering is
dominant.

\section{Results and Discussion}

\subsection{Cloaking of a sphere}
Aiming at the validation of our method, we design the susceptibility of the cover of the silicon dioxide ($\chi_1 = 2.9$) sphere shown in
Fig. \ref{fig:Figure_1}(b) with $\rho = r/R = 0.8$ to achieve transparency under a $x$-polarized electric field. This problem admits an analytical
solution (see for instance \cite{Bohren1998,Alu05}) that we use to estimate the error of our approach. We first solve this problem in the
small-particle limit, then, by using the radiation corrections presented in the appendix, we extend the solution to the case of
particle's sizes comparable to the wavelength.

\begin{figure}
\centering
\includegraphics[width=\mylength]{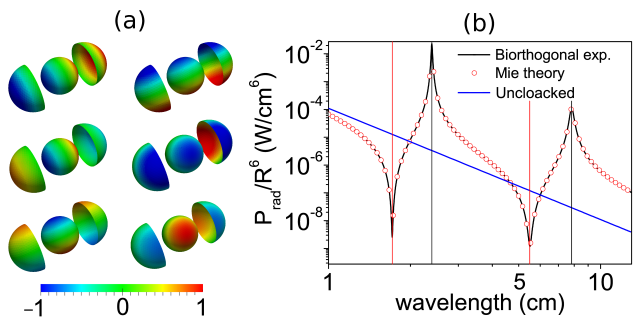}
 \caption{(a) Electric charge density of the eigenmodes of the coated sphere with non-vanishing radiative strength. The outer shell is
``opened'' to
allow the visualization of the surface charge density on the inner surface. (b) Power scattered by the coated sphere calculated with the
bi-orthogonal expansion and with the Mie-Theory. $R$ is the external radius
(see Fig \ref{fig:Figure_1} (b)). The power scattered by the
uncloaked Si sphere is also shown (blue line). The zeros and the poles of the x-component of the total  dipole moment are shown with red  and
black  vertical lines.} 
\label{fig:Figure_2}
\end{figure}

We start by numerically solving the electrostatic eigenvalue problems \ref{eq:EigProblem_Sigma} and \ref{eq:EigProblem_Tau}, obtaining the set of
resonant susceptibilities $\psi_k$ and the corresponding eigenmodes $\left( \sigmaonk, \sigmatwk \right) $ and $\left( \tauonk, \tautwk \right)$,
normalized according to Eq.  \ref{eq:Nomalization}
\cite{SM_numerical}. Then, the values of $\rkinconz$ and $\rkinctwz$ and of the
resonant radiative strength ${\bf s}_k$ are calculated for each mode using Eqs. \ref{eq:rk_zero_order} and \ref{eq:RadiativeStrenght}. At this
point, we
notice that only two degenerate eigenvalues (each of them with multiplicity 3) are associated to eigenmodes with non-vanishing radiative strength.
Their surface charge density is shown in Fig. \ref{fig:Figure_2} (a) and their resonant susceptibilities are listed in Tab. \ref{tab1}.
Furthermore, for symmetry considerations the total dipole moment has to be oriented along the $x$-axis.
%% 236
%
The coupling between these six eigenmodes, formally described by Eq. \ref{eq:DesignFormula} with $n=6$ and $\hat{\alpha} = \hat{\bf x}$, gives rise
to zeros in the scattered power. The values of
susceptibility $\chizero$ satisfying the transparency condition, i.e. Eq. \ref{eq:DesignFormula}, are listed in Tab. \ref{tab1};   they  are real
and in very good agreement with the analytical solution (error below 0.1\%). To achieve transparency at a given wavelength $\lambda_0$,
the  susceptibility of the cover should satisfy the constraint $\chi_2 \left( \lambda_0 \right) = \chizero $. 
It
is worth noting that since actual materials always exhibit losses, the transparency condition is never exactly satisfied. If a Drude
metal is the material of choice, its plasma  frequency $\omega_p$ is given by:
\begin{table}
\begin{tabular}{ccc}
\toprule
 {\# mode } & 1,2,3 & 4,5,6 \\
\colrule
$\psi_k$  & -13.9  & -1.3 \\
\colrule
{ \# zero }   & 1 & 2 \\
\colrule
$\chizero \text{ (theory)} $  & -0.6665 & -6.8474 \\
$\chizero \text{ (numeric) } $   & -0.6663 &  -6.8545 \\
\colrule
\botrule
\end{tabular}
\caption{Resonant susceptibilities of the modes of a spherical core-shell objects ($\chi_1=2.9, \, \rho=0.8$) with non-vanishing resonant radiative
strength and zeros of the x-component of the overall dipole moment of the coated sphere.}
\label{tab1}
\end{table}
%% WC  22
\begin{equation}
\omega_p =  \sqrt{- \left( \omega_0^2+\gamma^2 \right) \chizero }.
 \label{eq:PlasmaFrequency}
\end{equation}
%% WC 110
By assuming $\lambda_0 = 5.5 cm$ and  $\gamma=8 \cdot 10 ^{8} s^{-1}$ and choosing the zero $\# 2$,  i.e. $\chizero =  -6.8545$,
we obtain $\omega_p=8.97\cdot 10 ^{10} s^{-1}$ rad/s.
For instance, a micro-structure made of a regular array of thin wires can be properly designed to mimic a Drude metal with the prescribed plasma
frequency \cite{Pendry96}.

In Fig. \ref{fig:Figure_2} (b), we plot the corresponding scattered power as a function of the wavelength obtained using Eqs. \ref{eq:DipolarTotal}
and \ref{eq:RadiatedPower} where the summation runs  only over the modes $1 \ldots 6$. We also show with a blue line the scattered power from the
uncloaked sphere and with red open circles the scattered power of the cloaked sphere calculated using the Mie theory, which validates the
bi-orthogonal expansion.

\begin{figure}[ht]
\centering
\includegraphics[width=\mylength]{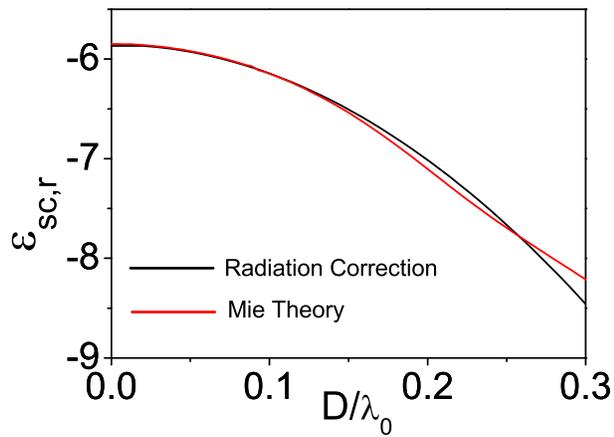}
\caption{Scattering-cancellation permittivity $\epsr$  of the coating of a spherical object ($\chi_1=2.9, \, \rho=0.8$)  computed by using the 
radiation corrections (black line) and the full-wave Mie theory (red line) as a function of the electric size of
the object $D/\lambda_0$. }
  \label{fig:Figure_4bis}
\end{figure}

As soon as the diameter $D$ of the sphere becomes comparable to the incident wavelength, the quasi-static prediction of the scattering-cancellation
permittivity made by Eq. \ref{eq:DesignFormula} becomes inaccurate and the use of the radiation corrections presented in 
the appendix is mandatory. Thus, we determine the radiation corrections to the zero $\#2$ ($\epszeror =  -5.8545$) as a function of
the electric size
$D/\lambda_0$ of the sphere and we compare the resulting value of $\epsr$ with the one obtained by the Mie theory using Ref. \cite{Alu05}. By using
Eqs.
\ref{eq:RadiationCorrection1} and \ref{eq:RadiationCorrection2} we obtain $\epsoner = - 4.1018 \cdot 10^{-06} j$ and $\epstwor = -0.7332$. It is
worth noting that the value of $\epsoner$ is negligible, thus $\epsr$ can be approximated by: 
\begin{equation}
  \epsr \approx \epszeror + \epstwor \left(2 \pi \frac{D}{\lambda_0}\right)^2 
  \label{eq:Correction2order}
\end{equation}
In Fig. \ref{fig:Figure_4bis} we plot the scattering-cancellation permittivity calculated by Eq. \ref{eq:Correction2order} and by the Mie
theory \cite{Alu05}.
We notice very good agreement between the two approaches up to the electric size of $D/\lambda_0=0.3$.

\begin{figure}[ht]
\centering
\includegraphics[width=\mylength]{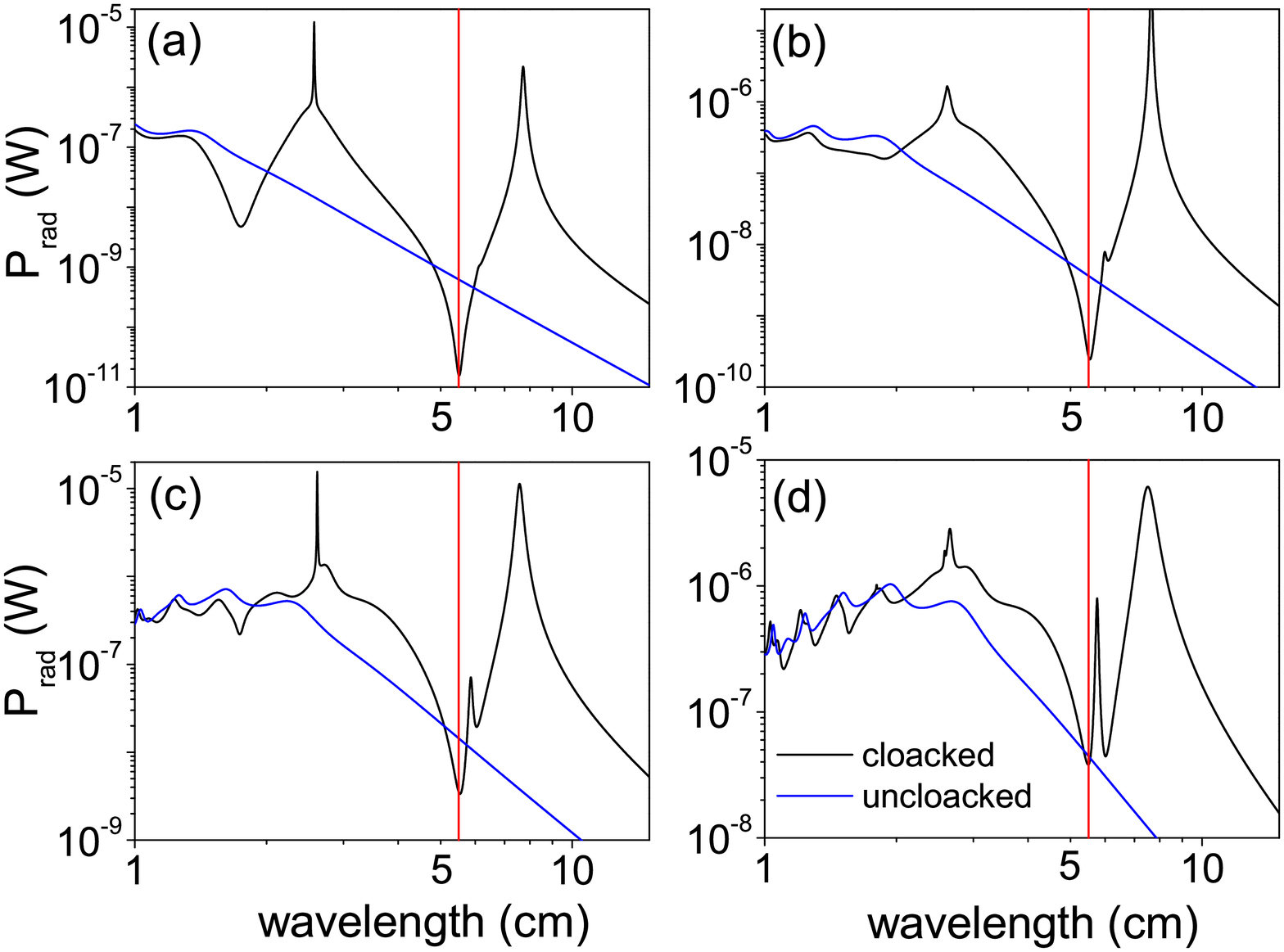}
\caption{Spectra of the power scattered by the investigated coated sphere  ($\chi_1=2.9, \, \rho=0.8$) for different electric sizes, namely $D/\lambda
= 0.15$ (a)  ,$0.2$ (b) , $0.25$(c) , $0.3$ (d) calculated by the full-wave Mie theory. The material of the coating has been designed using the
quasi-electrostatic theory and the radiation corrections. The used values of plasma frequency of the coating are listed in Tab \ref{tab_sphere_2}. The
power scattered by the uncloaked Si sphere (blue line) and the nominal wavelength $\lambda_0 = 5.5cm$ (red line) are also shown.}
  \label{fig:Rad_Sphere}
\end{figure}

\begin{table}[ht]
\begin{tabular}{ccccc}
\toprule
 { $ D/\lambda_0$ }   & 0.15 & 0.2 & 0.25 & 0.3 \\
\colrule
${\epsr}$   &  -6.5  &  -7.0&-7.7 &  - 8.5  \\
${\omega_p} (rad/s)$   & $9.39 \cdot 10^{10}$  & $ 9.70 \cdot 10^{10}$ & $1.01 \cdot 10^{11}$ & $ 1.05 \cdot 10^{11}$  \\
${Q}$  &  16dB &  11.5dB & 6.3dB &  0.67dB  \\
\botrule
\end{tabular}
\caption{ For different electric sizes of the coated-sphere we list the permittivity ${\epsr}$ of the coating at which scattering-cancellation
occurs, the corresponding value of plasma frequency ${\omega_p}$ of the Drude cover and the achieved quality of scattering-cancellation $Q$ defined
as the ratio between the powers scattered  by the uncloaked and the cloaked object at $\lambda_0=5.5$cm.}
 \label{tab_sphere_2}
\end{table}
Choosing $\lambda_0 = 5.5cm$ and four different electric sizes, namely $D/\lambda_0 = 0.15,0.2,0.25,0.3$ we obtain the corresponding values of $\epsr$
that guarantee the dipole scattering-cancellation using Eq. \ref{eq:Correction2order}. They are listed in Tab. \ref{tab_sphere_2}
together with the corresponding plasma frequencies of the Drude-metal coating obtained using Eq. \ref{eq:PlasmaFrequency}. Thus, for each case, we
plot in Fig.
\ref{fig:Rad_Sphere} the power scattered by the coated sphere as a function of the wavelength. In all the four cases the minimum of the scattering
power  corresponds  to the nominal wavelength $\lambda_0=550nm$ (vertical red line) as expected.
Nevertheless, we notice a degradation of the quality of scattering-cancellation as the electric size of the  object increases  due to the 
onset of
higher-order scattering modes. In particular, in correspondence to $D/\lambda_0=0.3$ (Fig. \ref{fig:Rad_Sphere} (d)), despite the scattered power
has
still a local minimum at $\lambda_0$, the scattered powers of the cloaked and the uncloaked structure are almost equal. For larger sizes, 
the cancellation of the dipole-scattering is not sufficient to reduce the total scattered power, since the scattering is dominated by higher order
modes.

\subsection{Cloaking of a C-shaped object}

%% 412	
In order to demonstrate the feasibility of the presented approach, we design the material of the cover of the C-shaped object sketched in Fig.
\ref{fig:Figure_1} (c-d) to achieve transparency under a $y$-polarized excitation. The core is made of silicon dioxide, i.e. $\chi_1=2.9$, with
dimensions $R
=
0.3,\, d=0.3,\, a= 0.5 $, while the thickness of the cover is $\delta=0.05$.  We numerically solve the eigenvalue problems \ref{eq:EigProblem_Sigma}
and
\ref{eq:EigProblem_Tau}, obtaining the set of resonant susceptibilities $\psi_k$ and the corresponding
eigenmodes $\left( \sigmaonk, \sigmatwk \right) $ and $\left( \tauonk, \tautwk \right)$. Then, the values of $\rkinconz$ and $ \rkinctwz $ and of the
resonant radiative strengths ${\bf s}_k$ are calculated for each mode using Eqs. \ref{eq:rk_zero_order} and \ref{eq:RadiativeStrenght}. 
\begin{figure}
\centering
\includegraphics[width=\mylength]{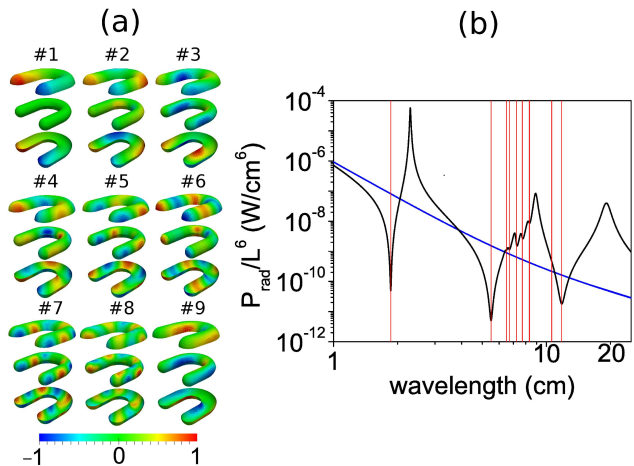}
 \caption{(a) Electric charge density of the eigenmodes of the coated C-shaped particle with appreciable radiative strengths when excited with a
$y$-polarized electric field. The outer shell is ``opened'' to allow the visualization of the surface charge density on the inner surface. (b) Power
scattered by a coated C-shaped particle when excited by a $y$-polarized electric field. $L$ is the horizontal length of the particle (see Fig.
\ref{fig:Figure_1}(c)). The power scattered by the uncloaked C-shaped particle is also shown (blue line). The zeros of the y-component of the total
dipole moment are also shown with red vertical lines.}
  \label{fig:Figure_4}
\end{figure} 
At this point, we notice that only nine eigenvalues are associated to eigenmodes with non-negligible radiative strength. The remaining eigenmodes
have resonant radiative strengths less than a prescribed limit $\| {\bf s}_k \| /  \max_k \| {\bf s}_k \| < 1.2\cdot 10^{-4}$ and have been
disregarded. The
surface charge
density of the eigenmodes is shown in Fig. \ref{fig:Figure_4} (a) and their resonant susceptibilities $\psi_k$ are reported in table \ref{tab3}. Thus,
we
assume that the component of the dipole moment along the incident polarization direction is dominant. The
values of
susceptibility satisfying the transparency condition, i.e. Eq. \ref{eq:DesignFormula} with $n=9$ and with $\hat{\alpha} = \hat{\bf y}$, are listed in
Tab. \ref{tab3}. To achieve transparency at a given wavelength $\lambda_0$, the susceptibility of the cover should satisfy the constraint $\chi_2
\left( \lambda_0
\right) = \chizero$, where $\chizero$ is the generic root of Eq. \ref{eq:DesignFormula}.
Since actual materials always exhibit losses, when  a zero of
the dipole moment is in  proximity of a pole, i.e. a plasmon resonance, the pole-zero cancellation can deteriorate the quality of the designed
transparency. Therefore, in the presence of real materials the roots of Eq. \ref{eq:DesignFormula} are not
equivalent in
terms of the quality of the scattering-cancellation. When many solutions are allowed, as in the present scenario, the zeros far from the poles have to
be preferred. In this case, examining Tab. \ref{tab3}, we select the zero $\# 8$, i.e. $\chizero = -7.4$.
Considering a Drude metal,  and assuming $\lambda_0 = 5.5cm$ and
$\gamma= \omega_p  \cdot 10 ^{-2} $ we obtain $\omega_p =9.29 \cdot 10^{10}$ rad/s. At this point, the condition \ref{eq:Hypothesis} has been verified
{\it a posteriori} using Eq. \ref{eq:DipolarTotal}. In Fig. \ref{fig:Figure_4} (b) we plot the corresponding scattered power
as a function of the wavelength using Eqs. \ref{eq:DipolarTotal} and \ref{eq:RadiatedPower} where the summation runs only over the modes $1
\ldots 9$ \cite{SM_validation}. We also show with a blue line the power  scattered by the uncloaked C-shaped object and with red
vertical
lines the position of the zeros of the y-component of the overall dipole moment, which are directly obtained from the values of susceptibility. At the
wavelength $\lambda_0 = 5.5cm$ the scattered power is reduced of $24.8dB$ with respect to the uncloaked object.  It is worth noting that in Fig.
\ref{fig:Figure_4} (b) the zeros of the scattered power spectra give rise to asymmetric scattering
line-shapes, usually referred to as Fano-like resonances. As already shown for arrays of homogeneous plasmonic objects \cite{ForestierePRB_13}, also
in strongly subwavelength plasmonic shells Fano-like resonances are originated by dipole scattering-cancellation of bright-modes
\cite{Argyropoulos12}.

\begin{table}[ht]
\begin{tabular}{cccccccccc}
\toprule
 { \# mode }   & 1 & 2 & 3 & 4 & 5 & 6 & 7 & 8 & 9 \\
\colrule
${\psi_k}$   & -89.0  &  -27.3 & -19.2 &  -16.5 &  -14.2 & -12.4 &  -10.7 &  -10.1 &   -1.3 \\
${\lambda_k}$ (cm) &  19.1 &  10.6 &  8.9 &   8.2 &   7.6 &   7.1 &   6.6  &   6.5 &  2.3 \\
\colrule
 { \# zero }   & 1 & 2 & 3 & 4 & 5 & 6 & 7 & 8 & 9 \\
\colrule
 ${\chizero}$ & -33.8  & -27.3 & -16.8 & -14.5 & -12.7 &  -10.8 & -10.2 &  -7.4 &  -0.84 \\
  $\lambdaSC$ (cm) & 11.8 & 10.6 &8.3 &   7.7  &  7.2 &   6.7  &   6.5 &    5.5 &    1.9 \\
\colrule
\botrule
\end{tabular}
\caption{Resonant susceptibilities of the bright modes and zeros of the $y-$component of the total dipole moment of a coated C-shaped object excited
by a $y$-polarized electric field.  The corresponding wavelengths are also listed assuming $\omega_p =9.29 \cdot 10^{10}$  rad/s and $\gamma= \omega_p
\cdot 10 ^{-2}$.}
 \label{tab3}
\end{table}

As the diameter $D$ of the minimum sphere circumscribing the C-shaped object become comparable to the incident wavelength $\lambda_0$, the
scattering-cancellation permittivity calculated by Eq. \ref{eq:DesignFormula} has to be corrected using the perturbation approach. Thus, we determine
the radiation corrections to the zero $\#8$ of Tab. \ref{tab3} ($\epszeror = -6.4$) as a function of the electric size $D/\lambda_0$ of the
C-shaped object. By using Eqs. \ref{eq:RadiationCorrection1} and \ref{eq:RadiationCorrection2} we obtain $\epsoner = - 4.5105 \cdot 10^{-05} j$ and
$\epstwor = -0.0763$. Since the value of $\epsoner$ is negligible, also in this case $\epsr$ can be approximated by Eq.  \ref{eq:Correction2order}.
In
Fig. \ref{fig:Figure_4bis} we plot the scattering-cancellation permittivity calculated by Eq. \ref{eq:Correction2order} as a function of the electric
size of the object $D/\lambda_0$.

\begin{figure}[ht]
\centering
\includegraphics[width=\mylength]{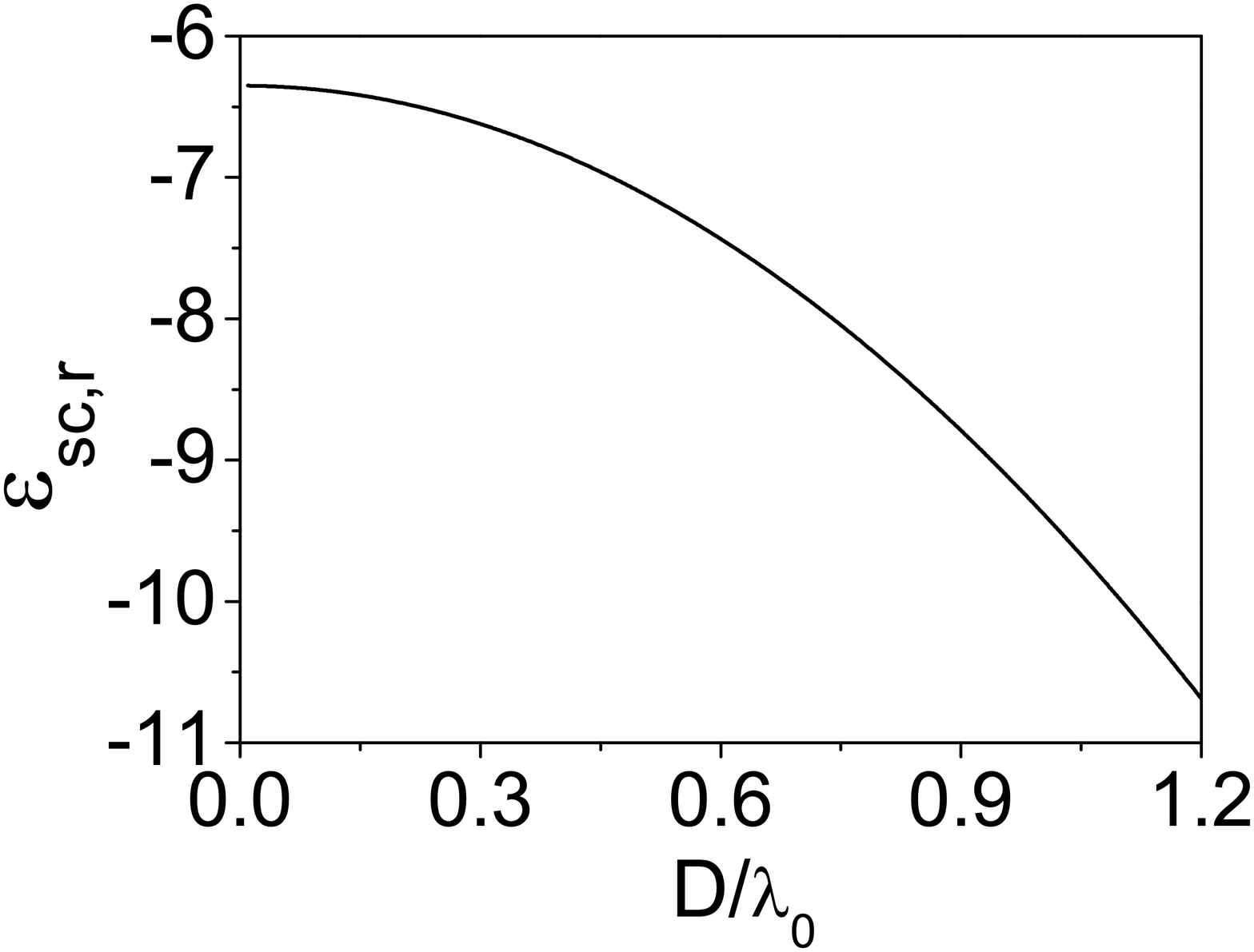}
\caption{Scattering-cancellation permittivity $\epsr$ of the coating of a C-shaped object as a function of the electric size
of the object $D/\lambda_0$ and computed using the radiation corrections.}
  \label{fig:Figure_5bis}
\end{figure}

\begin{figure}[ht]
\centering
\includegraphics[width=\mylength]{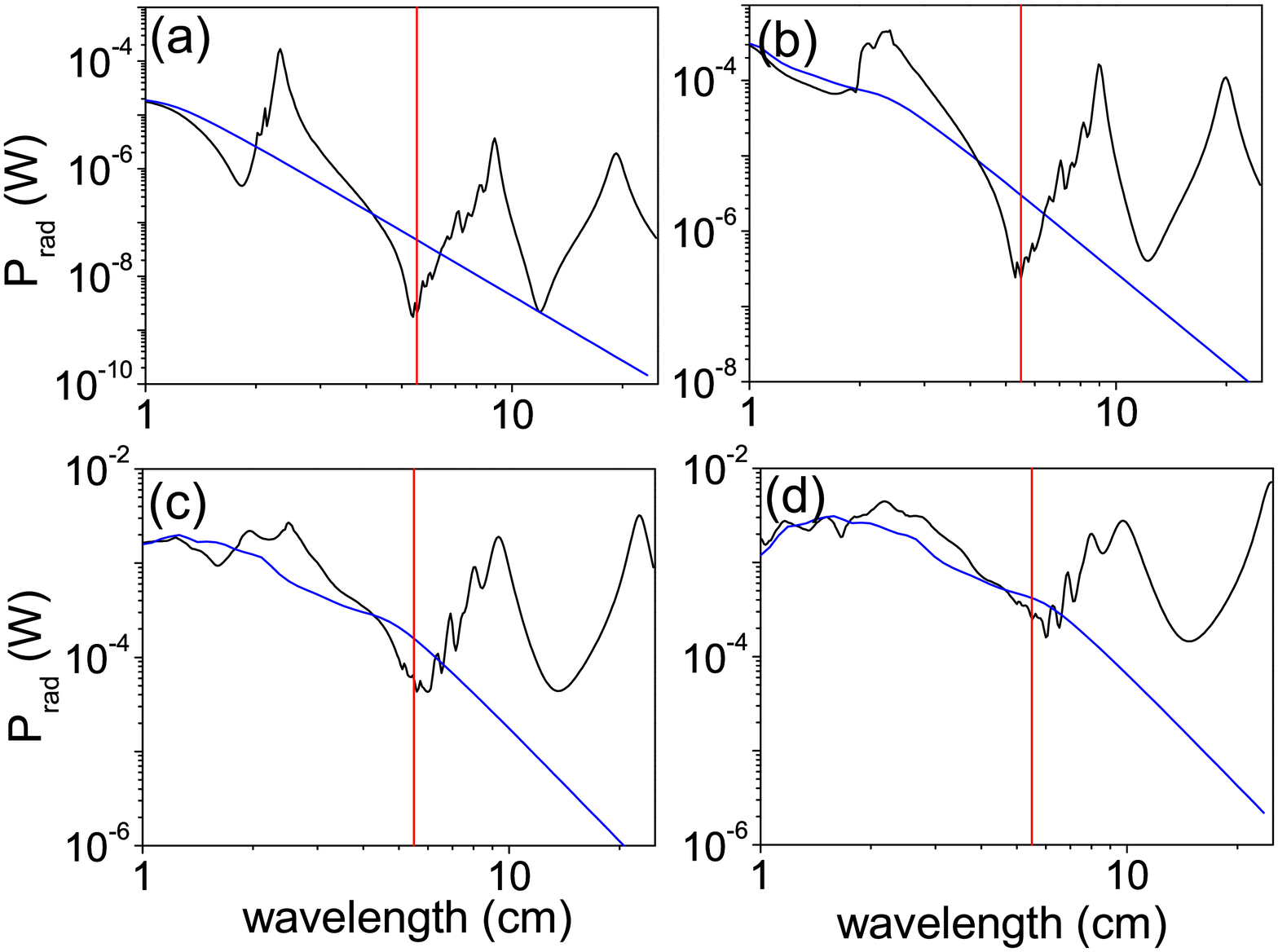}
\caption{Spectra of the power scattered by the investigated coated C-shaped particle for different electric sizes, namely $D/\lambda = 0.2$(a),
 $0.6$(b), $0.8$(c), $1$(d), when it is excited by a y-polarized plane wave propagating along $z$. The spectra have been calculated by the
full-retarded Method of Moments \cite{HarringtonBookMoM,Yla-Oijala05b}. The material of the coating has been designed using the quasi-electrostatic
theory and the radiation corrections. The used values
of plasma frequency of the coating are listed in Tab \ref{tab_C_2}. The power scattered by the uncloaked C-shaped particle (blue line) and the 
wavelength $\lambda_0 = 5.5cm$ (red line) are also shown.}
  \label{fig:Figure_5tr}
\end{figure}

By choosing $\lambda_0 = 5.5cm$ and four different electric sizes, namely $D/\lambda_0 = 0.18,0.54,0.72,0.9$, we obtain the values
of $\epsr$, listed in Tab. \ref{tab_C_2}, that guarantee the dipole scattering-cancellation. Thus, for each value of $\epsr$ we show in
Fig. \ref{fig:Figure_5tr} the power scattered by the coated C-shaped object as a function of the wavelength. In all the four cases the minimum of
the scattering power corresponds to the nominal wavelength $\lambda_0=5.5cm$ (vertical red line). This fact validates our design.
 As
in the case of
the sphere, the onset of higher-order scattering modes, occurring for large electric sizes has a detrimental effect on quality of
scattering-cancellation. Nevertheless, it is worth noting that a moderate reduction of the scattered power is achieved also in the case
in which the dimension of the C-shaped object is equal to the incident wavelength $\lambda_0$ as shown Fig. \ref{fig:Figure_5tr}.

\begin{table}[ht]
\begin{tabular}{ccccc}
\toprule
 { $ D/\lambda_0$ }   & 0.2 & 0.6 & 0.8 & 1 \\
\colrule
${\epsr}$   &   -6.45  & -7.23	&  -7.91 &   -8.79  \\
${\omega_p} (rad/s)$   & $9.35 \cdot 10^{10}$  & $ 9.83 \cdot 10^{10}$ & $1.023 \cdot 10^{11}$ & $ 1.072 \cdot 10^{11}$  \\
${Q}$  &  13dB &  7.3dB & 4.3dB &  2.2dB  \\
\botrule
\end{tabular}
\caption{ For different electric sizes of the coated C-shaped object we list the permittivity ${\epsr}$ of the coating at which
scattering-cancellation occurs, the corresponding value of plasma frequency ${\omega_p}$ of the Drude cover and the achieved quality of scattering
cancellation $Q$ defined as the ratio between the powers scattered  by the uncloaked and the cloaked object at $\lambda_0=5.5$cm.}
 \label{tab_C_2}
\end{table}

\section{Conclusions}
%% WC 169
We have introduced a novel theory for the cloaking of arbitrarily-shaped objects through designed homogeneous coatings.  Our approach, 
which is valid beyond the Rayleigh regime, permits the rigorous design of the permittivity values of the coating  layer that abates the total
scattered
power.  It can be also easily extended to design the cloaking of objects lying on a substrate, 
multi-coated objects, and plasmonic 
cores with a dielectric shell.

Nevertheless, it is important to point out that the achieved cloaking depends on the polarization of the incident light, being this limitation
inherent to the scattering-cancellation approach to cloaking when applied to arbitrary shapes using homogeneous and isotropic materials
\cite{SM_polarization}.  Moreover, since this approach is based on the cancellation of the dipole scattering, it is ineffective when the
electric size of the particle is large enough that high orders of scattering are dominant. Moreover, in the presence of objects of extreme aspect
ratio it may not be possible to find a direction $\hat{\boldsymbol \alpha}$ satisfying the condition \ref{eq:Hypothesis}. In this case the scattering
cancellation is not limited by the losses but by a residual polarization lying on the plane orthogonal to $\hat{\boldsymbol \alpha}$ with a consequent
degradation of the quality of the
cloaking \cite{SM_hypothesis}.

Despite its limitations, the introduced framework paves the way to the application of the scattering-cancellation to real-life
problems where the shape of the object to be cloaked, e.g. a complex sensor, is not under the control of the designer. 
\acknowledgments
This work was supported by the U.S. Army Research Laboratory through the Collaborative Research Alliance (CRA) for MultiScale
multidisciplinary Modeling of Electronic materials (MSME), and by the Italian Ministry
of Education, University and Research through the project PON01\_02782.
\appendix*
\section{Derivation of the radiation correction of the scattering-cancellation permittivity}
\label{sec:RadiativeCorrection}
In the present section we derive the first- and second-order radiation correction of the permittivity at which scattering-cancellation occurs.
This is achieved by zeroing the corresponding  perturbations of the  $\hat{\boldsymbol \alpha}$-component  of the dipole moment of the
coated object. In order to accomplish this, we have adapted to the problem at hand the approach that  Mayergoyz et al. originally introduced to study
the radiation correction of the plasmonic resonance \cite{Mayergoyz05}.
Thus, by introducing the normalized incident and scattered fields
\begin{equation}
    \begin{aligned}
        {\bf e}_{inc} &= \sqrt{\varepsilon_0} \, {\bf E}_{inc} \\ 
        {\bf h}_{inc} &= \sqrt{\mu_0} \, {\bf H}_{inc}
    \end{aligned}, \qquad
    \begin{aligned}
        {\bf e}_{t} &= \sqrt{\varepsilon_0} \, {\bf E}_{t} \\ 
        {\bf h}_{t} &= \sqrt{\mu_0} \, {\bf H}_{t}
    \end{aligned}
    \; \forall t \in \left\{1,2,3\right\} ,
\end{equation}
and scaling the spatial coordinates by the diameter $D$ of the smallest sphere circumscribing the object, we obtain the following boundary value
problem:
\begin{equation}
\begin{aligned}
        \nabla \times {\bf e}_1 & = -j \beta {\bf h}_1 \\
        \nabla \times {\bf h}_1 & = +j \beta \varepsilon_{1,r} {\bf e}_1 +j \beta \left( \varepsilon_{1,r} - 1
\right) {\bf e}_{inc} \\
        \nabla \cdot {\bf e}_1 & = {\bf 0}  \\ 
        \nabla \cdot {\bf h}_1 & = {\bf 0}  
     \end{aligned}
     \quad \mbox{in} \, V_1,
     \label{eq:ME_V1}
\end{equation}

\begin{equation}
\begin{aligned}
        \nabla \times {\bf e}_2 & = -j \beta {\bf h}_2\\
        \nabla \times {\bf h}_2 & = +j \beta \varepsilon_{2,r} {\bf e}_2 +j \beta \left( \varepsilon_{2,r} - 1
\right) {\bf e}_{inc} \\
        \nabla \cdot {\bf e}_2 & = {\bf 0}  \\ 
        \nabla \cdot {\bf h}_2 & = {\bf 0}  
     \end{aligned}
     \quad \mbox{in} \, V_2,
     \label{eq:ME_V2}
\end{equation}

\begin{equation}
\begin{aligned}
        \nabla \times {\bf e}_3 & = -j \beta {\bf h}_3 \\
        \nabla \times {\bf h}_3 & = +j \beta {\bf e}_3 \\
        \nabla \cdot {\bf e}_3& = {\bf 0}  \\ 
        \nabla \cdot {\bf h}_3& = {\bf 0} 
\end{aligned}
     \quad \mbox{in}  \, V_3,
     \label{eq:ME_V3}
\end{equation}

\begin{equation}
     \begin{aligned}
          {\bf n} \cdot \left( \varepsilon_{2,r} {\bf e}_2 - \varepsilon_{1,r} {\bf e}_1 \right) &=  - \left( \varepsilon_{2,r}  - \varepsilon_{1,r}
\right) {\bf n} \cdot  {\bf
e}_{inc}  \\
          {\bf n} \times \left( {\bf e}_2  - {\bf e}_1  \right) &= {\bf 0} \\
          {\bf n} \cdot \left( {\bf h}_2  - {\bf h}_1  \right) &= {\bf 0} \\
          {\bf n} \times \left( {\bf h}_2  - {\bf h}_1  \right) &= {\bf 0} 
     \end{aligned}
     \quad \mbox{on}  \, S_1,
     \label{eq:BC_S1}
\end{equation}

\begin{equation}
     \begin{aligned}
          {\bf n} \cdot \left(  {\bf e}_3 - \varepsilon_{2,r} {\bf e}_2 \right) &=  - \left(1  - \varepsilon_{2,r}
\right){\bf n} \cdot  {\bf
e}_{inc}  \\
          {\bf n} \times \left( {\bf e}_3  - {\bf e}_2  \right) &= {\bf 0} \\
          {\bf n} \cdot \left( {\bf h}_3  - {\bf h}_2  \right) &= {\bf 0} \\
          {\bf n} \times \left( {\bf h}_3  - {\bf h}_2  \right) &= {\bf 0} 
     \end{aligned}
     \quad \mbox{on}  \, S_2,
    \label{eq:BC_S2}
\end{equation}
where we have defined the quantity $\beta = \omega \sqrt{\epsilon_0 \mu_0} D$.

When the dimension $D$ is small compared to the free-space wavelength, the forcing terms ${\bf e}_{inc}, {\bf h}_{inc}$, the relative permittivity
${\varepsilon_{\mathtt{SC},r}}$ at which the scattering-cancellation occurs, and the corresponding fields ${\bf e}$ and ${\bf h}$ can be expanded in
powers of $\beta$, namely
\begin{equation}
   \begin{aligned}
      {\bf e}_{inc} & \approx \Einczero + \beta \Eincone + \beta^2 \Einctwo +  \cdots \\
      {\varepsilon_{\mathtt{SC},r}} & \approx {\epszeror} + \beta \epsoner +\beta^2 \epstwor + \cdots \\
      {\bf e} & \approx \Ezero + \beta \Eone + \beta^2 \Etwo +  \cdots \\
      {\bf h} & \approx \Hzero + \beta \Hone + \beta^2 \Htwo + \cdots \\
    \end{aligned}
    \label{eq:Expansion}
\end{equation}
In the particular case of a plane wave excitation of expression:
\begin{equation}
   \begin{aligned}
        {\bf e}_{inc} &= {\bf e}_0 \exp \left(- j \beta {\bf i}_k \cdot {\bf r} \right),
    \end{aligned}
\end{equation}
we have:
\begin{equation}
   \begin{aligned}
    \Einczero & = {\bf e}_0, \\
      \Eincone &=  - j \left( {\bf i}_k \cdot {\bf r} \right){\bf e}_0, \\
\Einctwo &= - \frac{\left( {\bf i}_k \cdot {\bf r}\right) ^2}{2} {\bf e}_0 .
\end{aligned}
\end{equation}
\subsection{Zero-Order Boundary Value Problem}
Substituting the expansion \ref{eq:Expansion}  in Eqs. \ref{eq:ME_V1}-\ref{eq:BC_S2} and equating the terms of zero-power we obtain the zero-order
boundary value problem for the electric field:
\begin{equation}
     \left\{
     \begin{aligned}
        \nabla \times \Ezero_t & = {\bf 0} \\ 
        \nabla \cdot \Ezero_t & = { 0} 
     \end{aligned} \domain
     \right. ,
     \label{eq:ME_E_zero_ord}
\end{equation}
\begin{equation} 
\begin{aligned}
          {\bf n} \cdot \left( \epszeror \Ezero_2 - \varepsilon_{1,r} \Ezero_1 \right) &=  - \left( \epszeror
 -
\varepsilon_{1,r} \right) {\bf n} \cdot \Einczero, \\ 
          {\bf n} \cdot \left(  \Ezero_3 - \epszeror \Ezero_2 \right) &=  - \left( 1
 -
\epszeror \right) {\bf n} \cdot \Einczero, \\
{\bf n} \times \left( \Ezero_2 - \Ezero_1 \right) &= {\bf 0}, \\
{\bf n} \times \left( \Ezero_3 - \Ezero_2 \right) &= {\bf 0}, 
\end{aligned}
   \label{eq:BC_E_zero_ord}
\end{equation}
and the zero-order boundary value problem for the magnetic field:
\begin{equation}
     \left\{
     \begin{aligned}
        \nabla \times \Hzero_t & = {\bf 0}, \\ 
        \nabla \cdot \Hzero_t & = { 0}, 
     \end{aligned} \domain
     \right.
     \label{eq:ME_H_zero_ord}
\end{equation}
\begin{equation}
\begin{aligned}
   {\bf n} \cdot \left(  \Hzero_2 - \Hzero_1 \right) &= 0, &
   {\bf n} \cdot \left(  \Hzero_3 - \Hzero_2 \right) &= 0, \\
   {\bf n} \times \left( \Hzero_2 - \Hzero_1 \right) &= {\bf 0}, &
   {\bf n} \times \left( \Hzero_3 - \Hzero_2 \right) &= {\bf 0}.
   \label{eq:BC_H_zero_ord} 
\end{aligned}
\end{equation}
First, from Eqs. \ref{eq:ME_H_zero_ord} and \ref{eq:BC_H_zero_ord} we can conclude that  $\Hzero=0$ in $\mathbb	{R}^3$. Next, we notice that the set
of Eqs. \ref{eq:ME_E_zero_ord}-\ref{eq:BC_E_zero_ord} defines the electrostatic problem encountered in the previous section. Thus, we have to use
Eq. \ref{eq:DesignFormula} in order to find the zero-order values $\epszeror$ of the dielectric permittivity zeroing the  $\hat{\boldsymbol
\alpha}$-component of the zero-order dipole
moment.
\subsection{First-Order Boundary Value Problem}
Next, equating the terms of first-power in
Eqs. \ref{eq:ME_V1}-\ref{eq:BC_S2} we obtain the first-order boundary value problem for the electric field:
\begin{equation}
     \begin{aligned}
        \nabla \times \Eone_t & = {\bf 0} \\
        \nabla \cdot \Eone_t & = { 0} 
     \end{aligned} 
   \domain,
  \label{eq:ME_E_1}
\end{equation}
\begin{equation}
     \begin{aligned}
          {\bf n} \cdot \left( \epszeror \Eone_2 - \varepsilon_{1,r} \Eone_1 \right) =& - \epsoner {\bf n} \cdot
\left(\Ezero_2 + \Einczero \right) \\ & - \left( \epszeror  -  \varepsilon_{1,r} \right) {\bf n} \cdot  \Eincone,  \\
          {\bf n} \cdot \left( \Eone_3 - \epszeror \Eone_2 \right) =& + \epsoner {\bf n} \cdot
\left(\Ezero_2 + \Einczero \right) \\ & - \left( 1 - \epszeror \right) {\bf n} \cdot  \Eincone ,  \\
           {\bf n} \times \left( \Eone_2 - \Eone_1 \right) &=  {\bf 0}, \\
           {\bf n} \times \left( \Eone_3 - \Eone_2 \right) &=  {\bf 0} .
     \label{eq:BC_E_1}
     \end{aligned}
\end{equation}
and the first-order boundary value problem for the magnetic field:
\begin{equation}
     \begin{aligned}
       \nabla \times \Hone_1 & = j {\varepsilon_{1,r}} \Ezero_1 + j \left(
{\varepsilon_{1,r}} - 1 \right) \Einczero, \\
       \nabla \times \Hone_2 & = j {\epszeror} \Ezero_2 + j \left(
{\epszeror} - 1 \right) \Einczero, \\ 
       \nabla \times \Hone_3 & = j  \Ezero_3, \\
       \nabla \cdot \Hone_t & = { 0} \domain, 
     \end{aligned}
     \label{eq:ME_H_1}
\end{equation}
\begin{equation}
     \begin{aligned}
          {\bf n} \cdot \left( \Hone_2 -  \Hone_1 \right) &= {\bf 0} &
          {\bf n} \cdot \left( \Hone_3 -  \Hone_2 \right) &= {\bf 0}, \\
          {\bf n} \times \left( \Hone_2 - \Hone_1 \right) &=  {\bf 0} &
          {\bf n} \times \left( \Hone_3 - \Hone_2 \right) &=  {\bf 0}. \\
     \end{aligned}
     \label{eq:BC_H_1}
\end{equation}
From Eqs. \ref{eq:ME_E_1}-\ref{eq:BC_E_1} we  derive the expression of the first-order correction $ {\bf p}^{\left( 1 \right)}$ of the
dipole moment of the coated object in terms of its electrostatic modes ${\bf p}_k$:
\begin{multline} 
   {\bf p}^{\left( 1 \right)} \left( \epsoner \right) = \\
    \displaystyle\sum_k  \frac{ \epsoner \left( \rkstwz + \rkinctwz \right)  + \chi_1 \rkincono + \chizero \rkinctwo }{  \psi_k -
\chizero} {\bf p}_k,
\label{eq:DipolarMoment_One}
\end{multline}
where we have defined the quantities:
\begin{equation}
 \begin{aligned}
  \rkincono  &= - {2\varepsilon_0}  \oint_{S_1}  \tauonk  {\bf n} \cdot \Eincone  dS,  \\ 
  \rkinctwo  &=   {2\varepsilon_0} \left(  \oint_{S_1} \tauonk {\bf n} \cdot \Eincone  dS  -
\oint_{S_2} \tautwk    {\bf n} \cdot \Eincone dS \right),  \\
  \rkstwz  &=   {2\varepsilon_0} \left(  \oint_{S_1} \tauonk {\bf n} \cdot \Ezero_1  dS  -
\oint_{S_2} \tautwk    {\bf n} \cdot \Ezero_2 dS \right).
  \end{aligned}  
   \label{eq:rk_first_order}
\end{equation}
By zeroing the quantity $ {\bf p}^{\left( 1 \right)}$ in Eq. \ref{eq:DipolarMoment_One} we obtain the first-order correction $\epsoner$ of the
permittivity  at which scattering-cancellation occurs: 
\begin{equation}
  \epsoner = - \frac{\displaystyle\sum_k  \frac{ \chi_1 \rkincono + \chizero \rkinctwo}{{  \psi_k - 
\chizero}}p_{k,  \hat{\alpha}} }{\displaystyle\sum_k  \frac{ \rkstwz + \rkinctwz }{  \psi_k - 
\chizero }p_{k,  \hat{\alpha}} }.
  \label{eq:RadiationCorrection1}
\end{equation}
We now summarize the algorithm for the computation of  $ \epsoner$.
Assuming that the zero-order scattering cancellation permittivity $\epszeror = \chizero -1 $ has been already obtained using Eq.
\ref{eq:DesignFormula}, we calculate the corresponding scattered electric field $\Ezero$, solution of the
problem \ref{eq:ME_E_zero_ord}-\ref{eq:BC_E_zero_ord},  on $S_1$ and $S_2$. Thus, we apply Eq. \ref{eq:rk_first_order} to compute the quantities
$\rkstwz$ and $\rkincono, \rkinctwo$ using the fields $\Ezero$, $\Eincone$ and the electrostatic eigenvectors $\left( \tauonk, \tautwk \right)$.
Eventually, we calculate $\epsoner$ by Eq. \ref{eq:RadiationCorrection1}.

Let us now turn our attention to the first-order boundary problem for the magnetic field, shown in Eqs. \ref{eq:ME_H_1}-\ref{eq:BC_H_1}. Its solution
that will be useful for the second-order problem,  is given by:
\begin{multline}
   \Hone \left( Q \right) = j \frac{\epszeror - \varepsilon_{1,r}}{4\pi } \oint_{S_1} \frac{{\bf n}_M
\times \left( \Ezero + \Einczero \right)} {
r_{MQ}} dV_M  \\ + j \frac{ 1 - \epszeror}{4\pi } \oint_{S_2} \frac{{\bf n}_M
\times \left( \Ezero + \Einczero \right)} {
r_{MQ}} dV_M. 
\end{multline}

\subsection{Second-Order Boundary Value Problem}
Next, we discuss the second-order correction $\epstwor$ for the scattering-cancellation permittivity. Equating the terms of second-power in Eqs.
\ref{eq:ME_V1}-\ref{eq:BC_S2} we obtain
the second-order boundary value problem for the electric field:
\begin{equation}
       \begin{aligned}
           \nabla \times \Etwo_t & = -j \Hone_t \\
           \nabla \cdot \Etwo_t & = {\bf 0}
       \end{aligned}
       \quad 
       \mbox{in} \, V_{t}  \quad \forall i \in \left\{ 1, 2, 3 \right\} \, , 
\end{equation}
\begin{equation}
\begin{aligned}
          &{\bf n} \cdot \left( \epszeror \Etwo_2 - \varepsilon_{1,r} \Etwo_1 \right)  =  - \epstwor {\bf n}
\cdot \left(\Ezero_2 + \Einczero \right)  \\ & - \epsoner {\bf n}
\cdot
\left(\Eone_2 + \Eincone \right)   - \left( \epszeror - \varepsilon_{1,r} \right) {\bf n} \cdot  \Einctwo \, , \\
 &         {\bf n} \cdot \left(  \Etwo_3 - \epszeror \Etwo_2 \right) =   + \epstwor {\bf n}
\cdot
\left(\Ezero_2 + \Einczero \right)  \\ & +  \epsoner {\bf n}
\cdot
\left(\Eone_2 + \Eincone \right)   - \left(1 - \epszeror \right) {\bf n} \cdot  \Einctwo \, ,
\end{aligned}
\end{equation}
\begin{equation}
          \begin{aligned}
           {\bf n} \times \left( \Etwo_2 - \Etwo_1 \right) &=  {\bf 0} \, , \\
          {\bf n} \times \left( \Etwo_3 - \Etwo_2 \right) &=  {\bf 0} \, .
\end{aligned}
\end{equation}
As suggested in Ref. \cite{Mayergoyz05}, it is now convenient to split $ \Etwo_{t}$ into two components:
\begin{equation}
  \Etwo_{t} = \tEtwo_{t} + \ttEtwo_{t} \domain.
\end{equation}
The first component $\tEtwo_{t}$ is the solution of the following problem:
\begin{equation}
 \left\{
\begin{aligned}
        \nabla \times \tEtwo_t & =-j \Hone_t \\
        \nabla \cdot \tEtwo_t & = { 0} \\
     \end{aligned} 
     \right. \domain,
     \label{eq:ME_E_2_bar}
\end{equation}
\begin{equation}
     \begin{aligned}
         {\bf n} \cdot \left(  \tEtwo_2 -  \tEtwo_1 \right)  &= { 0}, &  {\bf n} \cdot \left(  \tEtwo_3 -  \tEtwo_2 \right)
 &= { 0}, \\
          {\bf n} \times \left( \tEtwo_2 - \tEtwo_1 \right) &=  {\bf 0}, & {\bf n} \times \left( \tEtwo_3 - \tEtwo_2 \right) &=  {\bf 0},
     \end{aligned}
     \label{eq:BC_E_2_bar} 
\end{equation}
while the second component $\ttEtwo_{t}$  satisfies the following boundary value problem:
\begin{equation}
 \left\{
\begin{aligned}
        \nabla \times \ttEtwo_t & = {\bf 0} \\
        \nabla \cdot \ttEtwo_t & = { 0} 
     \end{aligned} 
     \right. \domain
     \label{eq:ME_E_2_barbar},
\end{equation}
\begin{equation}
\begin{aligned}
         & {\bf n} \cdot \left( \epszeror \ttEtwo_2 - \varepsilon_{1,r} \ttEtwo_1 \right) =   - \epstwor {\bf n}
\cdot
\left(\Ezero_2 + \Einczero \right)  \\ &- \epsoner {\bf n}
\cdot
\left(\Eone_2 + \Eincone \right)    - \left( \epszeror - \varepsilon_{1,r} \right) {\bf n} \cdot \left(  \Einctwo + \tEtwo \right), \\
         & {\bf n} \cdot \left(  \ttEtwo_3 - \epszeror \ttEtwo_2 \right) =   \epstwor {\bf n}
\cdot
\left(\Ezero_2 + \Einczero \right)    \\ & + 	\epsoner {\bf n}
\cdot
\left(\Eone_2 + \Eincone \right)   + \left( \epszeror - 1   \right) {\bf n} \cdot \left(  \Einctwo + \tEtwo \right),
\end{aligned}
\label{eq:BC_E_2_barbar_n}
\end{equation}

\begin{equation}
          \begin{aligned}
           {\bf n} \times \left( \ttEtwo_2 - \ttEtwo_1 \right) &=  {\bf 0}, \\
          {\bf n} \times \left( \ttEtwo_3 - \ttEtwo_2 \right) &=  {\bf 0}. 
\end{aligned}
\label{eq:BC_E_2_barbar_t}
\end{equation}
By using the same line of reasoning of Ref. \cite{Mayergoyz05} we obtain the expression of $\tEtwo$ that satisfies Eqs. \ref{eq:ME_E_2_bar} and
\ref{eq:BC_E_2_bar}:
\begin{multline} 
    \tEtwo \left( P \right) = \\
 \frac{\left( \epszeror - \varepsilon_{1,r}  \right)}{8 \pi }   \oint_{S_1}   \frac{\left( {\bf
n}_M \times \left( \Ezero \left(M\right) + \Einczero \right) \right) \times{\bf r}_{MP}}{ r_{MP}}  dS_M + \\
 \frac{\left(  1 - \epszeror  \right)}{8 \pi }   \oint_{S_2}   \frac{\left( {\bf
n}_M \times \left( \Ezero \left(M\right) + \Einczero \right) \right) \times{\bf r}_{MP}}{ r_{MP}}  dS_M,
  \label{eq:barEformula}
\end{multline}
Then, from Eqs.  \ref{eq:ME_E_2_barbar}-\ref{eq:BC_E_2_barbar_t} we obtain the expression of the second-order correction $ {\bf p}^{\left( 2 \right)}$
of
the dipole moment of the coated-object in terms of its resonant modes ${\bf p}_k$:
\begin{widetext}
\begin{equation} 
   {\bf p}^{\left( 2 \right)} =  \displaystyle\sum_k  \frac{ \epstwor \left( \rkstwz + \rkinctwz \right) +
\epsoner \left( \rkstwo + \rkinctwo \right)  + \chi_1 \left( \rkincont + \rksont \right) + \chizero \left( \rkinctwt + \rkstwt
\right)
}{  \psi_k - 
\chizero} {\bf p}_k,
\label{eq:DipolarMoment_Eps}
\end{equation}
\end{widetext}
where we have defined the quantities:
\begin{equation}
 \begin{aligned}
  \rkincont  &= - {2\varepsilon_0}  \oint_{S_1}  \tauonk  {\bf n} \cdot \Einctwo  dS,  \\ 
  \rkinctwt  &=   {2\varepsilon_0} \left(  \oint_{S_1} \tauonk {\bf n} \cdot \Einctwo  dS  -
\oint_{S_2} \tautwk    {\bf n} \cdot \Einctwo dS \right),  \\
  \rkstwo  &=   {2\varepsilon_0} \left(  \oint_{S_1} \tauonk {\bf n} \cdot \Eone_2 dS  -
\oint_{S_2} \tautwk    {\bf n} \cdot \Eone_2 dS \right), \\
  \rksont  &= - {2\varepsilon_0}  \oint_{S_1}  \tauonk  {\bf n} \cdot \tEtwo  dS,  \\ 
  \rkstwt  &=   {2\varepsilon_0} \left(  \oint_{S_1} \tauonk {\bf n} \cdot \tEtwo  dS  -
\oint_{S_2} \tautwk    {\bf n} \cdot \tEtwo dS \right).  \\
  \end{aligned}  
   \label{eq:rk_second_order}
\end{equation}
By zeroing the quantity $ {\bf p}^{\left( 2 \right)}$ we  obtain the second-order correction of the permittivity $\epstwor$ at which
scattering-cancellation occurs: 
\begin{widetext}
\begin{equation}
 \epstwor = - \frac{ \displaystyle\sum_k  \frac{ \epsoner \left( \rkstwo + \rkinctwo \right)  + \chi_1 \left(
\rkincont + \rksont \right) + \chizero \left( \rkinctwt + \rkstwt \right)
}{  \psi_k - \chizero} p_{k,  \hat{\alpha}} }{ \displaystyle\sum_k  \frac{  \rkstwz + \rkinctwz
 }{  \psi_k - 
\chizero} p_{k,  \hat{\alpha}}  }.
\label{eq:RadiationCorrection2}
\end{equation}
\end{widetext}
We now recapitulate the steps for the computation of the second-order radiation correction $ \epstwor$.
Assuming that the calculation of $ \epsoner$ has been already performed, we calculate  the corresponding scattered electric field $\Eone$, solution
of the problem  \ref{eq:ME_E_1}-\ref{eq:BC_E_1}, on both the internal and the external surface. Moreover, using the electrostatic field
$\Ezero$ and the incident field $\Einczero$ we can calculate by Eq. \ref{eq:barEformula} the field $\tEtwo$ on the both internal and the external
surface. Thus, we compute the quantities $\rkincont, \rkinctwt, \rkstwo,\rksont,\rkstwt $. Eventually, $\epstwor$ can be calculated using Eq.
\ref{eq:RadiationCorrection2}.


\begin{thebibliography}{10}

\bibitem{Pendry06}
J.~B. Pendry, D.~Schurig, and D.~R. Smith, {\em Science}, vol.~312, no.~5781, pp.~1780--1782, 2006.

\bibitem{Leonhardt06}
U.~Leonhardt, {\em Science}, vol.~312, no.~5781,
  pp.~1777--1780, 2006.

\bibitem{Alu05}
A.~Al\`u and N.~Engheta, {\em Phys. Rev. E}, vol.~72, p.~016623, Jul 2005.

\bibitem{Silveirinha08}
M.~Silveirinha, A.~Al\`u and N.~Engheta, {\em Phys. Rev. B}, vol.~78, p.~205109, Nov 2008.

\bibitem{Schurig06}
D.~Schurig, J.~J. Mock, B.~J. Justice, S.~A. Cummer, J.~B. Pendry, A.~F. Starr,
  and D.~R. Smith,  {\em Science}, vol.~314, no.~5801, pp.~977--980, 2006.

\bibitem{Rainwater12}
D.~Rainwater, A.~Kerkhoff, K.~Melin, J.~C. Soric, G.~Moreno, and A.~Alù,
 {\em New Journal of Physics}, vol.~14, no.~1, p.~013054, 2012.

\bibitem{Alu09}
A.~Al\`u and N.~Engheta,  {\em Phys. Rev. Lett.},
  vol.~102, p.~233901, Jun 2009.

\bibitem{Tricarico2010}
S.~Tricarico, F.~Bilotti, A.~Al\`u, and L.~Vegni,  {\em Phys. Rev.
  E}, vol.~81, p.~026602, Feb 2010.

\bibitem{Ouyang89}
F.~Ouyang and M.~Isaacson, {\em Ultramicroscopy}, vol.~31,
  no.~4, pp.~345 -- 349, 1989.

\bibitem{Fredkin03}
D.~R. Fredkin and I.~D. Mayergoyz, {\em Phys. Rev. Lett.}, vol.~91, p.~253902, Dec
  2003.

\bibitem{Mayergoyz05}
I.~Mayergoyz, D.~Fredkin, and Z.~Zhang, {\em Phys. Rev. B}, vol.~72, p.~155412, 2005.

\bibitem{Mayergoyz07}
I.~Mayergoyz and Z.~Zhang, {\em Magnetics, IEEE Transactions on}, vol.~43, no.~4,
  pp.~1689--1692, 2007.

\bibitem{ForestierePRB_13}
C.~Forestiere, L.~Dal~Negro, and G.~Miano,  {\em Phys.
  Rev. B}, vol.~88, p.~155411, Oct 2013.

\bibitem{Hung13}
L.~Hung, S.~Y. Lee, O.~McGovern, O.~Rabin, and I.~Mayergoyz,  {\em Phys. Rev. B} {\bf 88}, 075424 (2013).

\bibitem{Borwein95}
P.~Borwein and T.~Erdelyi, {\em Polynomials and Polynomial Inequalities}.
\newblock New York: Springer-Verlag, 1995.

\bibitem{EngetaBook}
N.~Engheta and R.~W. Ziolkowski, {\em Electromagnetic Metamaterials: Physics
  and Engineering Explorations}.
\newblock New York: Wiley \& Sons, 2006.

\bibitem{Bohren1998}
C.~F. Bohren and D.~R. Huffman, {\em Absorption and Scattering of Light by
  Small Particles}.
\newblock Wiley, 1998.

\bibitem{SM_numerical}
``See supplemental material for greater details on the numerical solution.,''

\bibitem{Pendry96}
J.~B. Pendry, A.~J. Holden, W.~J. Stewart, and I.~Youngs, {\em Phys. Rev. Lett.},
  vol.~76, pp.~4773--4776, Jun 1996.

\bibitem{SM_validation}
``See supplemental material for the validation of the calculated scattered
  power using the method of moments({MoM}).,''

\bibitem{Argyropoulos12}
C.~Argyropoulos, P.-Y. Chen, F.~Monticone, G.~D'Aguanno, and A.~Al\`u,
  {\em Phys. Rev. Lett.}, vol.~108, p.~263905, Jun 2012.

\bibitem{HarringtonBookMoM}
R.~Harrington, {\em Field Computation by Moment Methods}.
\newblock New York: Macmillan, 1968.

\bibitem{Yla-Oijala05b}
P.~Yla-Oijala and M.~Taskinen, {\em Antennas and
  Propagation, IEEE Transactions on}, vol.~53, pp.~3316--3323, oct. 2005.

\bibitem{SM_polarization}
``See supplemental material for the investigation of the dependence of the
  transparency condition on the incident polarization.,''

\bibitem{SM_hypothesis}
``See supplemental material for an example in which this hypothesis is not
  completely satisfied.,''

\end{thebibliography}
\end{document}